\documentclass[conference]{IEEEtran}

\usepackage{graphicx}
\PassOptionsToPackage{hyphens}{url}\usepackage{hyperref}

\usepackage[inline]{enumitem}
\usepackage{pdfpages}

\ifCLASSOPTIONcompsoc
  \usepackage[nocompress]{cite}
\else
  \usepackage{cite}
\fi
\usepackage{url}

\makeatletter
\def\blfootnote{\xdef\@thefnmark{}\@footnotetext}
\makeatother

\IEEEoverridecommandlockouts
\IEEEpubid{\makebox[\columnwidth]{978-1-7281-8326-8/20/$31.00$~\copyright{}2020 IEEE \hfill}\hspace{\columnsep}\makebox[\columnwidth]{ }}

\begin{document}
\null%
\includepdf[pages={1}]{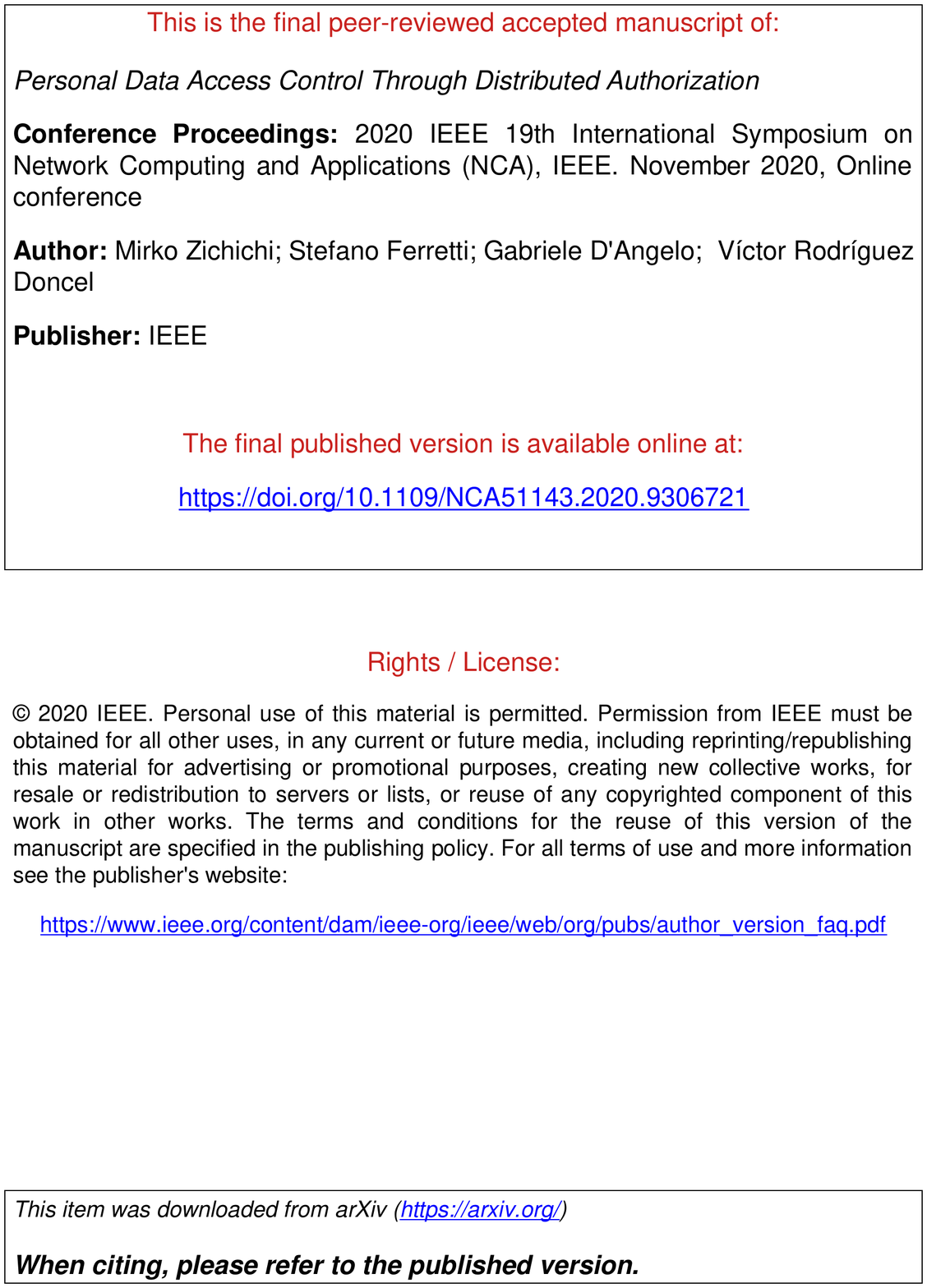}

\title{Personal Data Access Control Through\\Distributed Authorization}

\author{\IEEEauthorblockN{Mirko Zichichi\IEEEauthorrefmark{1}\IEEEauthorrefmark{2}\IEEEauthorrefmark{4},
Stefano Ferretti\IEEEauthorrefmark{3},
Gabriele D'Angelo\IEEEauthorrefmark{4}, 
Víctor Rodríguez-Doncel\IEEEauthorrefmark{2}}
\IEEEauthorblockA{\IEEEauthorrefmark{2}Ontology Engineering Group, Universidad Politécnica de Madrid, Spain\\ \emph{mirko.zichichi@upm.es, vrodriguez@fi.upm.es}}
\IEEEauthorblockA{\IEEEauthorrefmark{3}Department of Pure and Applied Sciences, University of Urbino ``Carlo Bo", Italy\\ \emph{stefano.ferretti@uniurb.it}}
\IEEEauthorblockA{\IEEEauthorrefmark{4}Department of Computer Science and Engineering, University of Bologna, Italy\\ \emph{g.dangelo@unibo.it}}}

\maketitle

\blfootnote{\IEEEauthorrefmark{1}This work has received funding from the European Union’s Horizon 2020 research and innovation programme under the Marie Skłodowska-Curie International Training Network European Joint Doctorate grant agreement No 814177 \href{https://www.last-jd-rioe.eu/}{Law, Science and Technology Joint Doctorate - RIoE}.}

\begin{abstract}
This paper presents an architecture of a Personal Information Management System, in which individuals can define the access to their personal data by means of smart contracts. These smart contracts, running on the Ethereum blockchain, implement access control lists and grant immutability, traceability and verifiability of the references to personal data, which is stored itself in a (possibly distributed) file system. A distributed authorization mechanism is devised, where trust from multiple network nodes is necessary to grant the access to the data. To this aim, two possible alternatives are described: a Secret Sharing scheme and Threshold Proxy Re-Encryption scheme. The performance of these alternatives is experimentally compared in terms of execution time. Threshold Proxy Re-Encryption appears to be faster in different scenarios, in particular when increasing message size, number of nodes and the threshold value, i.e.~number of nodes needed to grant the data disclosure.
\end{abstract}


%
\IEEEpeerreviewmaketitle

\section{Introduction} \label{sec:intro}
The transformation introduced by digital technologies has had (and is having) a significant impact on economy and society. Data is at the heart of this transformation and individuals are the main sources generating more and more of it. 
There is an urgent need to place (again) individuals at the center and to relieve the absence of technical instruments and standards that make the exercise of one's rights simple and not excessively burdensome \cite{davari2019access, zichichi2020efficiency}. 
The EU's GDPR \footnote{Council  of  European  Union, Regulation  2016/679  -  directive 95/46} helps to promote this vision and at the same time seeks to pave the way for open data spaces for the social and economic good \footnote{European Commission, COM(2020) 66, ``A European strategy for data''}.

Our aim is to seek such a technology by enabling users with the sovereignty over their data, while guaranteeing its confidentiality. In our view, the data owner can define access by limiting the scope of data utility, delegating these privileges or giving up ownership completely, without the need to rely on (un)trusted entities to facilitate this task.
The development of a Personal Information Management System (PIMS) \footnote{European  Data  Protection  Supervisors, Opinion 9/2016,  ``EDPS Opinion  on  Personal Information Management Systems''}  that fulfils these goals can be based on a distributed software architecture, where each individual is associated to a digital space containing personal data. This space will be used to attend the data access requests coming from data providers and data consumers.
Distributed Ledger Technologies (DLT) and Decentralized File Storages (DFS) combination provides a range of features suitable for data management and sharing, such as transparency, immutability and reliability \cite{zichichi2020efficiency, zichichi2020framework}. 

The contribution of the paper is the following. First, we propose an architecture for PIMS that, based on the use of DLTs, smart contracts and (D)FS enable to manage data access on the basis of multiple entities whose role is to provide mutual trust between the parties. Second, we employ two specific distributed authorization mechanisms, i.e.~Secret Sharing (SS) and Threshold Proxy Re-Encryption (PRE) that allow for a decentralized distribution of personal data, yet guaranteeing data sovereignity to users, confidentiality and a secure access control. Third, we present an experimental evaluation of these considered schemes, showing that PRE is faster in different scenarios, although it has the drawback of requiring that the user generates a re-encryption key each new data consumer.

The remainder of this paper is organized as follows. Section \ref{sec:back} presents the background concepts and Section \ref{sec:rel} describes related approaches. Section \ref{sec:das} specifies the system architecture. Performance is evaluated in Section \ref{sec:eval} before conclusions in Section \ref{sec:concl}.

\section{Background} \label{sec:back}
\subsection{Distributed Ledger Technologies and Smart Contracts}
DLTs provide a data ledger that guarantees the immutable persistence of the data, thereby ensuring untampered data availability. DLTs were born with the purpose of shifting trust from a human intermediary, who manages a transaction between two parties, to a protocol that allows two or more parties to conduct transactions directly. The resistance to manipulation makes DLTs able to support smart contracts.
These consist of an immutable set of instructions that are executed deterministically by several participants in a network, who receive the same inputs and then perform a calculation which leads to the same outputs. When the issuer of a smart contract broadcasts it on the DLT network -- and he is also sure that the implemented behaviour is correct (e.g. through code auditing) -- then the transactions originating from such a contract do not require the presence of a third party to be validated.
In Ethereum~\cite{buterin2013ethereum}, every process is completely traced and permanently stored in the blockchain because the smart contract computation is executed by all network participants. 

\subsection{Access Control Mechanisms}
The objective of access control systems is to regulate access to system resources by enforcing permissions on the basis of a set of system policies for determining who can access information.
Currently, most of them are based on a centralized controller that has the authority to access the data and, therefore, entails the risk of a single point of failure and, above all, privacy leakage \cite{jemel2017decentralized}. 
In this paper we study techniques that are usually not strictly related to access control mechanisms, but that can offer a better privacy guarantee.

\subsubsection{Secret Sharing}
A sophisticated cryptographic technique that allows providing privacy to users for their data consists in Secret Sharing (SS), firstly proposed by Shamir~\cite{shamir1979share} and Blakley~\cite{blakley1979safeguarding}. This $(t,n)$-threshold scheme is used to share a secret between a set of $n$ participants, with the possibility to reconstruct the secret using any subset of $t$ (with $t \leq n$) or more shares, but no subset of less than $t$.
By employing this in a network of (mostly honest) nodes, privacy is provided to a user that is sharing the secret, since none of the nodes can reconstruct the secret without the help of other $t-1$ nodes.

\subsubsection{Proxy Re-Encryption}
PRE \cite{ateniese2006improved}, is a type of public-key encryption, where a semi-trusted proxy entity transforms a ciphertext $c$, encrypted with a public key $pk_1$, into a ciphertext decryptable with a private key $sk_2$, without learning anything about the underlying plaintext. This is possible using a re-encryption key $rk_{1-2}$ generated by the data owner who has the key pair ($pk_1$, $sk_1$) and and that divulges (to the proxy) the authorization of access to the plaintext to a data consumer holding the keypair ($pk_2$, $sk_2$). Among many schemes, in single-use uni-directional proxy re-encryption, the re-encryption function is one way.

\section{Related Work} \label{sec:rel}
In the last years, many research efforts on access control have been done to securely store, share and transmit data while ensuring its integrity, validity and authenticity. Centralized controller issues may be addressed by adapting a solution based on a DLT for the verification of access permissions to an access control mechanism \cite{jemel2017decentralized}. 
However, when data volumes and sharing grows as fast as in online social networks or smart cities, it becomes difficult to manage access control and deal with personal data \cite{zichichi2020distributed, zichichi2020efficiency}.  
A possible approach would be to securely store access control policies on DLTs, whereby the applicant can be made aware of his or her permissions to access his or her personal data, as in \cite{zyskind2015decentralizing, yan2017bc}.
In particular, Yan et al. \cite{yan2017bc} employ the use of a SS scheme to share personal information in pieces between network nodes, however their innovative solution is expensive and not GDPR compliant due to the storing of personal data on the DLT.

Another possible approach is to program access control policies as smart contracts, in order to manage control automatically \cite{davari2019access}, and work on the breaking points between GDPR and DLTs. For this matter, Onik et al. \cite{onik2019privacy}, propose a model that stores personal data off-chain (i.e. not directly stored in the DLT) and traces its life cycle through data processors and processors by means of a DLT. 

\section{Decentralized Authorization Service} \label{sec:das}
\subsection{Data Storage System}
DLTs got momentum, mainly for their ability to provide immutability.
However, if we intend to comply with the GDPR, it requires the modification or deletion of data under certain circumstances, e.g. the ``right to be forgotten''.
Thus, in our system, personal data are stored \textit{encrypted} in an off-chain File Storage (FS) and then referenced in a DLT \cite{finck2019they}. This solution has the additional benefit of improving performances and to provide higher availability for data reads and writes \cite{zichichi2020efficiency}, and can be implemented either by using a DFS or a commercial cloud storage provider, since data is encrypted in the client device.
Once a file is published in the off-chain storage (1\textsuperscript{st} step in Figure \ref{fig:arch}), the returned reference can be employed to retrieve it and its digest allows the verification of its integrity \footnote{The full description of this data storage system is available in \cite{zichichi2020efficiency,zichichi2020framework} since the main focus, here, is on the authorization service.}. 
We implement pepper and salt in the hash function and when possible we try to consider a sufficiently large parameter space of the original data in order ‘not to allow the data subject to be identified via “all” “likely” and “reasonable” means’ \cite{AEPD2019hash,finck2019they}.

\subsection{Smart Contract Access Control}
\begin{figure}
    \centering
	\includegraphics[width=.48\textwidth]{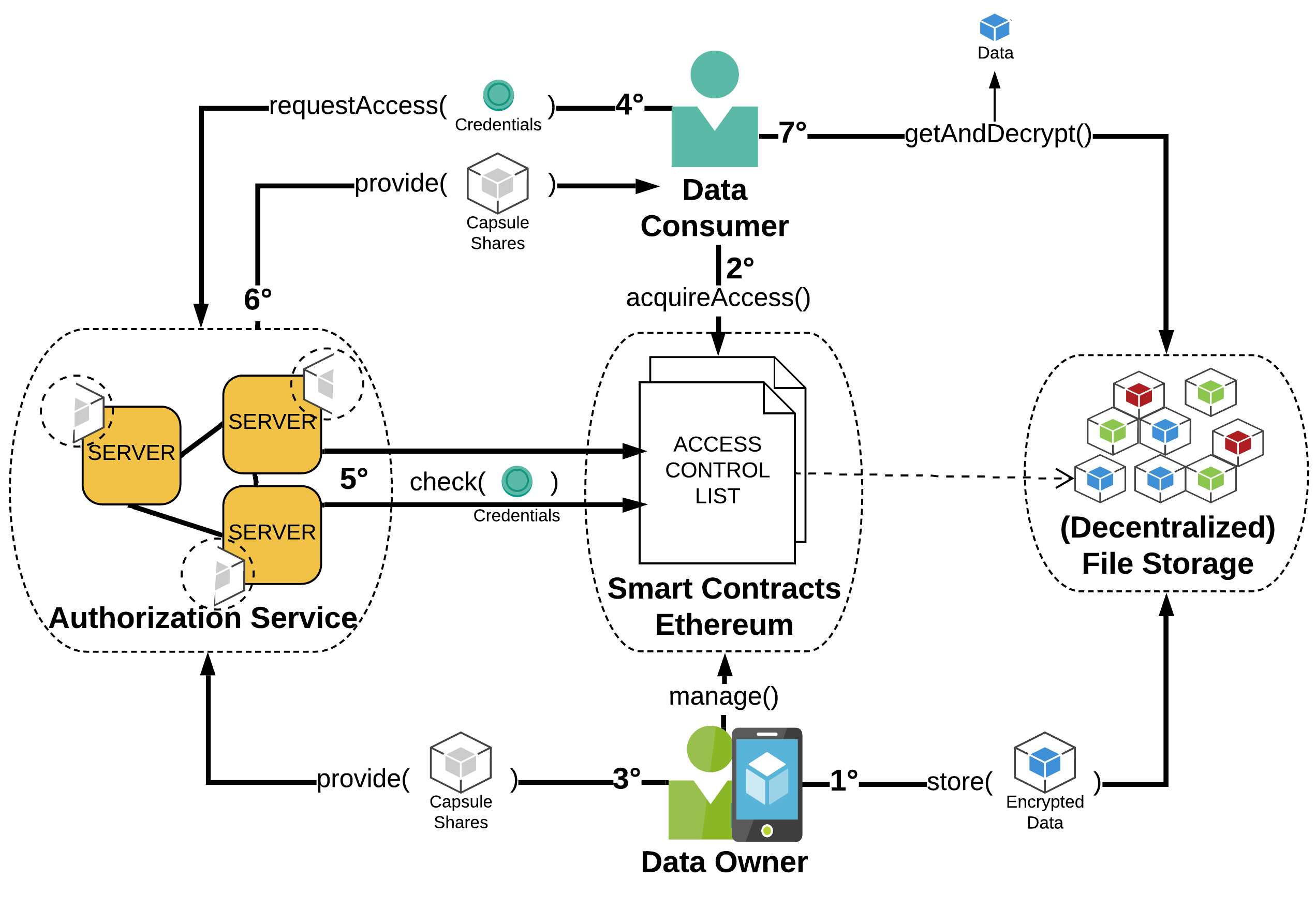}
	\caption{Architecture of the Decentralized Authorization Service}
	\label{fig:arch}
\end{figure}

Ethereum is the DLT where part of out the access control logic to share data is performed. Through smart contracts, access to the data can be purchased or can be allowed by the owner (2\textsuperscript{nd} step in Figure \ref{fig:arch}). The use of data, then, is authorized only to entitled users. Hence, due to the presence of smart contracts, no direct interactions are needed among the data owner and users interested in his data.
In practice, each piece of data is referenced in a specific smart contract in Ethereum.
The smart contract maintains an Access Control List (ACL) that represents the rights to access a bundle of data. 
Once a consumer is eligible to obtain certain data, i.e.~he is in the ACL, he can access such data through an access key, which is provided by the authorization service. 

\subsection{Cryptosystem}
We use a hybrid encryption scheme \cite{Herranz2006KEMDEMNA} that consists of an asymmetric public key part to encrypt a key, using a Key Encapsulation Mechanism (KEM), plus a symmetric secret key part to encrypt actual data through a Data Encapsulation Mechanism (DEM). This KEM/DEM technique leverages the combination of the efficiency and large message space of secret key cryptography with the benefits of public key cryptography \cite{Herranz2006KEMDEMNA}.
According to this approach, data generated from a data source device is encrypted using a symmetric content key $k_{DEM}$ in an efficient DEM. This key is stored in a ``capsule'' generated by a KEM, i.e.~it is encrypted using public key cryptography with a keypair ($pk_{KEM}$, $sk_{KEM}$).
In order to efficiently cope with time series data, we refer to Droplet’s key management \cite{shafagh2018droplet} to handle symmetric content keys in the DEM and on their use of Dual-Key Stealth Address Protocol to protect the privacy of sharing relationships in public DLTs.

\subsection{Authorization Service Network}
In our proposal, the authorization service is in charge of enforcing the access rights that are specified in the smart contracts ACLs. When this service is operated by a single central provider, trust must be given to this one, since the keys are kept in one place only. Assuming that this provider can be honest-but-curious, privacy may be threatened, e.g.~an online social network site sharing a user geolocation with his/her friends, if curious, can access to this information. Thus, we propose to decentralize the service in order to shift the trust to the protocol. In this case, nodes in a network are considered semi- or un-trusted, but a data protection/cryptographical mechanism, built into their execution protocol, allows the whole system to be trusted \cite{shamir1979share, blakley1979safeguarding, zyskind2015decentralizing, egorov2017nucypher}.

When a data consumer with keypair ($pk_c$,$sk_c$) is entitled to access some data in a smart contract ACL, he requests the release of the associated capsule to the authorization service through a message signed with $pk_c$, that proves that his address is in the ACL (4\textsuperscript{th} step in Figure \ref{fig:arch}).
Upon user request, the authorization service checks if the data consumer is eligible, through interaction with the smart contract (5\textsuperscript{th} step). If this is the case, i.e. the data consumer is on the ACL, the service starts the operation for releasing, or ``opening'', the capsule (6\textsuperscript{th} step) that holds the $k_{DEM}$ secret key needed to decrypt the desired data publicly stored in a (D)FS (7\textsuperscript{th} step).
We refer to two cryptographical schemes for the data owner's key management (3\textsuperscript{rd} and 6\textsuperscript{th} step):

\subsubsection{Secret Sharing (SS)} 
This scheme splits the $sk_{KEM}$ in $n$ shares, but only $t$ shares ($t<n$) are enough to "open" the capsule, i.e. decrypting the capsule in order to obtain $k_{DEM}$.
A $(t,n)$-threshold scheme is employed, thus, single nodes alone are unable to reconstruct $sk_{KEM}$, because they only save a portion of this key.

\subsubsection{Threshold Proxy Re-Encryption (PRE)} 
The capsule, initially obtained from the $pk_{KEM}$ by the data owner, can be re-encrypted by the service using a re-encryption key $pk_{O \rightarrow C}$ generated by the owner. The re-encrypted capsule, then, can be decrypted using $sk_c$ by the consumer to obtain the $k_{DEM}$ needed to decrypt the data. 
PRE usually involves only one semi-trusted proxy node, however, it can decide to not follow the conditional policies as instructed or it may collude with the consumer to attack the data owner’s private key. 
Instead of using a single re-encryption key, multiple proxies can be involved in a $(t,n)$-threshold scheme with ``re-encryption shares'', in such a way that these can be combined client-side by the data consumer. 

Both these techniques come with different advantages and disadvantages. SS relieves the user from any interaction during each key distribution, but at the same time if $t$ nodes are malicious then the user cannot intervene to stop the keys from getting leaked. On the other side, PRE has the drawback of requiring the user to generate a re-encryption key $pk_{O \rightarrow C}$ for each new consumer, however he has the option to stop producing new re-keys if some nodes are malicious.

\begin{figure*}[ht]
    \centering
	\includegraphics[width=\textwidth]{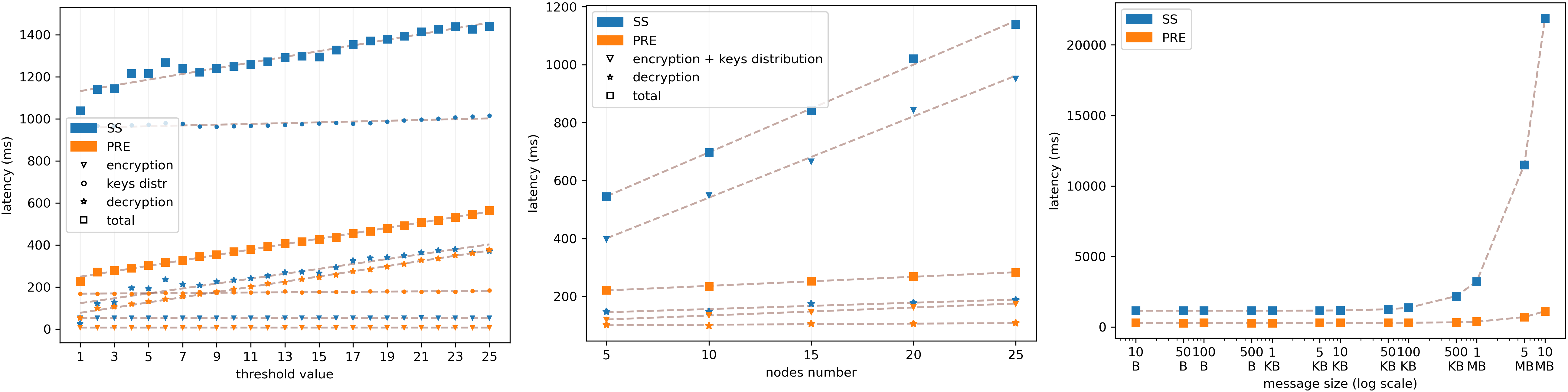}
	\caption{Encryption and decryption latencies while varying the threshold value, nodes number and messages size.}
	\label{fig:plots}
\end{figure*}
\section{Performance Evaluation} \label{sec:eval}
We measured the amount of time required to perform access control operations using an implementation of the proposed systems. We resort to the SS feature provided by the OpenEthereum client \cite{paritystore} and to the PRE implementation of NuCypher \cite{egorov2017nucypher}. The tests were performed using a network of 25 interconnected nodes with the aim to emulate the real DLTs and the distributed systems use cases \footnote{The reference software for tests can be found in Zenodo https://doi.org/10.5281/zenodo.4024735} 

We considered the use case where a data owner stores a smart contract containing an ACL in the Ethereum blockchain. Then, we emulated from 10 to 100 data consumers asking for access to some data after they have been added to the ACL. The results of the test carried out allow us to evaluate the goodness of the proposed approach in terms of performances:
\begin{itemize}
    \item \textbf{Threshold variation}: involves the variation of $t$ from $1$ to $25$, with number of nodes $n = 25$ and message size set to $30$ Bytes. As the first plot in Figure \ref{fig:plots} shows, the encryption time remains mostly constant ($\sim$7ms for PRE and $\sim$52ms for SS) while the decryption time increases linearly with $t$. The biggest time difference comes from the actual generation and distribution of the key shares in the encryption, $\sim$792ms, in favor of PRE. 
    \item \textbf{Number of nodes variation}: threshold value $t$ was set to $2$ and the message size was set to $30$ KB. Generally, as expected, the time costs of operations increase with the number of nodes $n$. However, we must note the fact that the slope of the curve in the second plot in Figure \ref{fig:plots} representing SS results is significantly higher than with PRE. This makes the PRE method more scalable.
    \item \textbf{Size of messages variation}: $n$ was set to $25$ and $t = 2$, while the size of the message varied. Results reported in the third plot Figure \ref{fig:plots} suggest that the PRE scheme scales beter than SS. From $10$ Bytes to $1$ MB PRE latency raises slightly, while SS has a clear inflection point when the message size is set to $100$ KB and then skyrockets from $1$ MB onward.
\end{itemize} 

\section{Discussion and Conclusion} \label{sec:concl}
In this paper, we presented the architecture of a PIMS, based on a decentralized approach for managing access to data where several entities have the role of providing mutual trust between the parties, i.e. the Authorization Service Network. We leveraged smart contracts to allow individuals to define access through an ACL to their personal data stored off-chain. 

We have focused on data protection through encryption, using two different schemes: SS and PRE. We showed their employment in the PIMS architecture and at first we discussed their qualitative differences, then we compared them in terms of execution time. Our performance evaluation shows that, in respect to SS, PRE is:
\begin{enumerate*}[label=(\roman*)]
    \item faster when increasing the size of the messages;
    \item more scalable, as it better manages the increase in the number of nodes executing the protocol;
    \item more efficient when increasing the threshold value, due to its shares generation method.
\end{enumerate*}
On the other hand, PRE has the drawback of requiring the data owner to generate a re-encryption key for each new data consumer. We also acknowledge the lack of scalability of the Ethereum blockchain and the expensiveness, in terms of cost per operation, needed to maintain a smart contract. 

In future work, we will pursue a more complex policy enforcement, adding another technological layer on top of the solution we presented, and we will compare it with more sophisticated methods, such as Attribute Based Encryption (ABE).



\bibliographystyle{IEEEtran}
\bibliography{paper}
%

\end{document}